\documentclass[11pt,a4paper]{article}
\usepackage[utf8]{inputenc}
\usepackage[T1]{fontenc}
\usepackage[a4paper,margin=2.5cm]{geometry}
\usepackage{amsmath,amssymb,amsfonts}
\usepackage{bm}
\usepackage{slashed}
\usepackage{graphicx}
\graphicspath{{./}{./figures/}}
\usepackage[numbers,sort&compress]{natbib}
\usepackage{url}
\usepackage[hidelinks,breaklinks]{hyperref}

\title{A Proposed $E_8 \times E_8$ Kinematic Scaffolding for the Standard Model with Pre-Gravitation}

\author{%
Priyank Kaushik,$^{1}$\quad Vatsalya Vaibhav,$^{2}$\quad Tejinder P. Singh$^{3}$\\[6pt]
\normalsize $^{1}$Indian Institute of Technology, Kanpur 208016, India\\
\normalsize $^{2}$Indian Institute of Technology, Kanpur 208016, India\\
\normalsize $^{3}$Tata Institute of Fundamental Research, Homi Bhabha Road, Mumbai 400005, India\\[4pt]
\normalsize\texttt{priyankk@iitk.ac.in}\quad\texttt{vatsalya@iitk.ac.in}\quad\texttt{tpsingh@tifr.res.in}%
}

\date{}

\begin{document}

\maketitle

\begin{abstract}
\noindent We present a proposed $E_8 \times E_8$ kinematic scaffolding for the standard model with pre-gravitation. The notation $E_8\times\omega E_8$ used below records a split-complex exchange grading of two factors; it is not a tensor product of groups. Each factor branches through $SU(3)\times E_6$ and thence through trinification, $E_6 \rightarrow SU(3)^3$. The first factor supplies representation labels associated with the standard-model lineage, and the second supplies a mirror, pre-gravitational lineage. The identification of a single vector-like colour group across the two factors, and the proposed gravitational reading of the right-sector $SU(2)$, remain dynamical hypotheses. Physical chiral fermions are not assigned to the $E_8$ adjoints: they are modelled by minimal ideals of the complex Clifford algebra $Cl_6(\mathbb C)$, while the $496$ adjoint dimensions are used only as a representation-label ledger. The numerical split $208+288$ is therefore a declared roster-matching convention, not an invariant decomposition and not a particle count. Spacetime enters through a selected six-dimensional split-biquaternionic vector space with a separately specified quadratic form of signature $(3,3)$. Its full frame group is $SO(3,3)$; deriving a soldering form and the proposed $BF$/Pleba\'nski dynamics remains open. The conventional Clifford--Dirac operator is constructed independently and exactly: $J_2(\mathbb H_s)$ realises the vector space $\mathbb R^{3,3}$ through its determinant, and after choosing $\mathbb H_s\cong M_2(\mathbb R)$ its matrix gradient and adjugate act between real four-dimensional Weyl modules and factorise the wave operator in both orderings. The rank-two algebra supplies the quadratic spacetime layer, whereas $J_3(\mathbb O_{\mathbb C})$ supplies the proposed internal, generation and cubic spectral layer through the magic-star decomposition of $\mathfrak e_8^{\mathbb C}$. We distinguish established algebraic identities from model identifications and list the missing real-form, anomaly, localisation and invariant-coupling constructions explicitly.
\end{abstract}

\vspace{1em}

\section{Motivation and overview}
\label{sec:intro}

There are strong reasons to seek a formulation of elementary particle physics which does not presuppose a classical spacetime manifold. Quantum theory as currently formulated needs a background time; the background is classical; and classicality is itself an emergent, approximate property of a quantum universe. The research programme to which this paper belongs \cite{TPsir, review, review2} takes as fundamental a pre-spacetime, pre-quantum matrix dynamics --- a generalised trace dynamics in the sense of Adler \cite{Adler_1994, Adler} --- whose configuration variables are matrices with entries valued in the division algebras and their split extensions. In such a theory the question ``what replaces spacetime and the gauge bundle?'' must be answered algebraically. The present paper records, in a single place, the group-theoretic scaffolding which the programme uses for that answer: the branching of $E_8 \times E_8$ through $SU(3)\times E_6$ and trinification, the pairing of the two factors by a split-complex tag, the resulting organisation of the standard-model and pre-gravitational sectors, and the construction of a six-dimensional spacetime arena of signature $(3,3)$ from the split biquaternions. Here $E_8\times\omega E_8$ is shorthand for an exchange-graded pair of $E_8$ sectors; it is not a tensor product with a number or a new Lie group.

Three structural commitments define the framework, and we state them at the outset because everything else in the paper is organised around them.

\emph{First: the physical chiral fermions are spinors, not adjoint components.} The fermions of one generation are realised as minimal left ideals of the complex Clifford algebra $Cl_6(\mathbb C)$ generated by octonionic multiplication chains \cite{Furey, Furey2018, Furey_2018, l-r}, with chirality supplied by a split bioctonionic doubling (Section~\ref{sec:fermions}). They are not components of an $E_8$ adjoint. The Distler--Garibaldi analysis \cite{DG} rules out the relevant class of attempts to obtain the observed chiral standard-model spectrum by embedding the Lorentz and gauge groups and the fermions directly in a real or complex form of $E_8$. The present construction avoids that class by placing the fermions in a separate Clifford module while using $E_8\times E_8$ as a ledger of representation labels (Section~\ref{sec:branching}). We make no claim that $E_8\times E_8$ already acts on the physical fermion module; constructing an invariant coupling is open problem (O1).

\emph{Second: spacetime is emergent, and its germ is split-algebraic.} The two $E_8$ factors are paired as $E_8 \times \omega E_8$, where $\omega$ is the split-complex unit, $\omega^2 = +1$, $\tilde\omega = -\omega$. The same $\omega$, acting one level down on the quaternionic triples, equips the eight-dimensional algebra of split biquaternions with a quadratic form of signature $(3,3)$ (Section~\ref{sec:arena}). The non-compact spacetime symmetry does not, and cannot, arise from the compact internal groups; it arises as the frame group $SO(3,3)$ of this split-graded quadratic space. That frame group is then gauged as a topological $BF$ theory whose constrained symmetry breaking yields Einstein gravity on one four-dimensional leaf and the weak interaction, as the geometry of a second, flipped-signature leaf \cite{Wesley, universeGTD}. The algebra automorphisms supply one diagonal compact rotation group; the boosts, and the remaining compact rotations, are frame transformations, supplied by the quadratic form, not by any algebra automorphism.

\emph{Third: the abelian charges are Clifford-algebraic, and the two $SU(2)$s with their hypercharges descend from trinification.} Electric charge is one third of the number operator of $Cl(6)$ \cite{Furey2018}; hypercharge is the derived combination $Y = Q - T^3_L$. The trinification chain $E_6 \rightarrow SU(3)^{\otimes 3}$, applied once per $E_8$ factor, delivers $SU(2)_L \otimes U(1)_Y$ in the visible sector and $SU(2)_R \otimes U(1)_{Y_{dem}}$ in the pre-gravitational sector. The frame sector supplies no abelian gauge charge beyond the $SO(2)$ that rotates the two discarded directions of the six-dimensional arena \cite{Wesley}; the hypercharges are internal, and trinification is where they live. The relation between the two non-abelian descriptions must be stated with equal care, and we state it here once: the ledger's trinification chain supplies the $SU(2)$ \emph{labels} and, with them, the hypercharges; the frame sector supplies the $SU(2)$ \emph{dynamics}; and the identification of the trinification $SU(2)_L$ and $SU(2)_R$ with the two chiral halves of the frame sector --- the precise sense in which the weak interaction is a spacetime symmetry in internal disguise --- is the programme's central structural conjecture, not a derived result. It is folded into open problems (O1) and (O3), and every later statement that leans on it should be read with that grade attached.

The paper is organised as follows. Section~\ref{sec:fermions} summarises the spinorial realisation of the fermions and fixes the ontology. Section~\ref{sec:branching} presents the complete $E_8\times E_8$ branching and the label ledger, including the matching rule, the $208+288$ arithmetic, and the status of the residual sector. Section~\ref{sec:generations} summarises the proposed origin of three generations in triality and the exceptional Jordan algebra, and the associated mass-ratio results. Section~\ref{sec:arena} constructs the six-dimensional arena, its two four-dimensional leaves, its gauging, and the associated gradient (Dirac-type) operator, with the mathematical status of each step stated explicitly. Section~\ref{sec:dynamics} describes the dynamical setting and its current epistemic status. Section~\ref{sec:anomalies} gives the anomaly structure. Section~\ref{sec:open} lists the open problems. Section~\ref{sec:outlook} gives the experimental outlook.

Throughout, we distinguish three grades of statement: \emph{derived} (a theorem or a standard result), \emph{conditional} (follows from stated hypotheses which are themselves plausible but unproven), and \emph{open} (a well-posed problem the programme has not solved). The grade is stated wherever it is not obvious from context.

\section{The physical fermions: spinors of the octonionic Clifford algebra}
\label{sec:fermions}

\subsection{One generation from $Cl(6)$}
\label{sec:cl6}

The starting point, which goes back to G\"unaydin and G\"ursey \cite{GunaydinGursey} and has been developed in recent years by Furey \cite{Furey, Furey2018, Furey_2018}, by Stoica \cite{Stoica}, and by others (see \cite{Baez} for the octonionic background and \cite{Dixon} for the division-algebraic programme at large), is the following. Left multiplication of the complexified octonions $\mathbb{C}\otimes\mathbb{O}$ on themselves generates, through nested (chained) left-multiplication maps, the complex Clifford algebra $Cl(6) \cong M_8(\mathbb{C})$. Fixing a unit imaginary octonion (conventionally $e_7$) equips the six remaining imaginary directions with three fermionic ladder operators $\alpha_1, \alpha_2, \alpha_3$ and their adjoints, obeying $\{\alpha_i, \alpha_j^\dagger\} = \delta_{ij}$, $\{\alpha_i,\alpha_j\}=0$.

A minimal left ideal of $Cl(6)$ is then a Fock space built on the idempotent projector to the Fock vacuum: it has complex dimension $8$, and its states, graded by the number operator
\begin{equation}
N_o = \sum_{i=1}^{3}\alpha_i^\dagger \alpha_i\,, \qquad N_o \in \{0,1,2,3\},
\label{eq:numberop}
\end{equation}
carry electric charge
\begin{equation}
Q = \frac{N_o}{3}\,,
\label{eq:charge}
\end{equation}
with spectrum $\{0, \tfrac13, \tfrac23, 1\}$. The eight states of one ideal are one neutrino, three anti-down quarks, three up quarks, and one positron; the conjugate ideal carries the eight antiparticle states (one antineutrino, three down quarks, three anti-up quarks, one electron). Charge is quantised in units of one third because a number operator has an integer spectrum \cite{Furey2018}; no anomaly-cancellation argument, grand-unified embedding, or monopole is invoked. The unitary symmetry preserving the ladder structure is $U(3) = (SU(3)\times U(1))/\mathbb{Z}_3$: the $SU(3)$ is colour, and the $U(1)$ generator is $N_o$, i.e.\ electric charge. Together, ideal and conjugate ideal supply sixteen states carrying precisely the internal quantum numbers of one standard-model generation with its right-handed (sterile) neutrino --- the fermion content of the $16$ of $Spin(10)$. We emphasise the arithmetic: one minimal ideal has dimension eight; the sixteen-dimensional generation content requires the ideal \emph{together with} its conjugate.

\subsection{Chirality from the split bioctonions}
\label{sec:chirality}

$Cl(6) \cong M_8(\mathbb{C})$ has a unique irreducible representation and is therefore structurally parity-blind: its ideals carry no handedness. Chirality requires a doubled algebra, and the split-complex unit supplies it. The split bioctonions $\mathbb{O}\oplus\omega\mathbb{O}$ generate the chain algebra $Cl(7) = Cl(6)\oplus Cl(6)$, whose two blocks are exchanged by the parity operation $\omega \rightarrow -\omega$ \cite{l-r}. One block houses the left-chiral states, the other the right-chiral states; each block carries a complete copy of the ladder system of Section~\ref{sec:cl6}. The left-handed states are electroweak doublets and the right-handed states are singlets of $SU(2)_L$, and mirror-wise for $SU(2)_R$; the explicit assignment, and the demonstration that $SU(2)_L$ and $SU(2)_R$ are parity-violating chiral gauge symmetries in this setting, are given in \cite{l-r} and, in the one-generation Clifford-algebraic setting, in \cite{NFurey}. (One caution, kept in force throughout: the gravi-weak reading developed in Sections~\ref{sec:arena} and \ref{sec:dynamics} reinterprets the right-sector factor geometrically, and the reconciliation of that reading with the internal chiral-gauge characterisation of $SU(2)_R$ in \cite{l-r} is part of open problem (O1).)

\subsection{The ontology, fixed}
\label{sec:ontology}

We can now state the fermion ontology of the framework in one sentence: \emph{the physical chiral fermions are the Clifford-module states of Sections~\ref{sec:cl6} and \ref{sec:chirality}; the $E_8\times E_8$ structure of the next section is an adjoint-lineage ledger which organises, but does not contain, those states.} This keeps the proposal outside the direct-embedding class analysed in \cite{DG}. It also creates a definite burden: an explicit $E_8\times E_8$-equivariant dynamics coupling the ledger, the Clifford module and the geometric sector has not been constructed. That is open problem (O1).

\section{The $E_8 \times E_8$ branching and the label ledger}
\label{sec:branching}

\subsection{Why $E_8 \times E_8$, and why paired by $\omega$}

The choice of $E_8\times E_8$ is motivated on three grounds. First, $E_8$ is the largest exceptional group, its adjoint is its smallest nontrivial representation, and its branching through $SU(3)\times E_6$ naturally produces both a trinification chain (housing the standard-model gauge lineage and a generation-counting $SU(3)$) and one further $SU(3)$ which we associate with the octonionic coordinate sector. Second, the doubling: the left--right (parity) structure of Section~\ref{sec:chirality} requires two mirror sectors, and the split-complex notation $E_8 \times \omega E_8$ tags the second factor by the same algebraic element $\omega$ used in the coordinate construction. Third, economy of coincidence: the exceptional chain $G_2 \subset F_4 \subset E_6 \subset E_7 \subset E_8$ is related to the octonions and the exceptional Jordan algebra $J_3(\mathbb{O})$ \cite{Ramond, truini}, structures also used in the fermion and generation ansatz. These parallels motivate the construction but do not derive a unique physical $E_8\times E_8$ action. We do not use $E_8\times E_8$ because string theory does, although the appearance of the same group in the heterotic theory \cite{PhysRevLett.54.502} is a suggestive point of contact (for heterotic standard-model constructions see \cite{Braun_2005}); the differences are noted where relevant.

\subsection{The branching chain}
\label{sec:chain}

Each $E_8$ branches as \cite{Slansky:1981yr}
\begin{equation}
     248 = (8,1) \oplus (1,78) \oplus (3,27) \oplus (\bar{3}, \overline{27})
     \label{eq1}
\end{equation}
under $SU(3)\times E_6$. After complexification, (\ref{eq1}) is the magic-star decomposition $\mathfrak{e}_8^{\mathbb C}=\mathfrak{sl}_3(\mathbb C)\oplus\mathfrak e_6^{\mathbb C}\oplus(\mathbf3\otimes\mathbf{27})\oplus(\bar{\mathbf3}\otimes\overline{\mathbf{27}})$ \cite{truini}, in which $\mathbf{27}$ can be realised as $J_3(\mathbb O_{\mathbb C})$ (Section~\ref{sec:generations}). The mixed sectors carry both indices and are therefore candidate algebraic channels for a geometry--internal coupling. Calling them a soldering mechanism would additionally require an explicit equivariant map to the tangent and matter bundles, which is not yet constructed. We name the leading $SU(3)$ of each factor $SU(3)_{spacetime,1}$ and $SU(3)_{spacetime,2}$. The $(8,1)$ comprises its generators, whereas the proposed octonionic coordinates transform as $1\oplus1\oplus3\oplus\bar3$ (Section~\ref{sec:coords}). The $E_6$ then branches through trinification, $E_6\rightarrow SU(3)^3$, with
\begin{equation}\label{eq2}
    27 = (\bar{3}, 3, 1) \oplus (3, 1, 3) \oplus (1, \bar{3}, \bar{3})
\end{equation}
\begin{equation}\label{eq3}
    \overline{27} = (3,\bar{3}, 1) \oplus (\bar{3}, 1, \bar{3}) \oplus (1, 3, 3)
\end{equation}
\begin{equation}\label{eq4}
    78 = (8,1,1) \oplus (1,8,1) \oplus (1,1,8) \oplus (3, 3, \bar{3}) \oplus (\bar{3}, \bar{3}, 3)
\end{equation}
In the first (visible) factor the three trinification $SU(3)$s are interpreted as $SU(3)_{genL}$ (generation labelling), $SU(3)_c$ (colour), and an $SU(3)$ which breaks to the electroweak $SU(2)_L \otimes U(1)_{\gamma_1}$; in the second (pre-gravitational) factor they are $SU(3)_{genR}$, the \emph{same} colour $SU(3)_c$ seen in the second octonionic frame (Section~\ref{sec:onecolour}), and an $SU(3)$ breaking to $SU(2)_R \otimes U(1)_{\gamma_2}$. The last $SU(3)$ branches with
\begin{equation}
       3 = 2(1) \oplus 1(-2), \qquad
       \bar{3} = 2(-1) \oplus 1(2), \label{eq6}
\end{equation}
\begin{equation}\label{eq7}
       8 = 1(0) \oplus 2(-3) \oplus 2(3) \oplus 3(0)
\end{equation}
where the integer in parentheses is the $U(1)_{\gamma}$ eigenvalue. Substituting into the trinification decompositions gives, for $E_6 \rightarrow SU(3)_{genL}\otimes SU(3)_c \otimes SU(2)_L \otimes U(1)_{\gamma_1}$,
\begin{equation}\label{eq8}
    27 = (\bar{3}, 3, 1)(0)  \oplus  (3, 1, 2)(1) \oplus (3, 1, 1)(-2)  \oplus (1, \bar{3}, 2)(-1) \oplus (1, \bar{3}, 1)(2)
\end{equation}
\begin{equation}\label{eq9}
         \overline{27} = (3, \bar{3}, 1)(0) \oplus (\bar{3}, 1, 2)(-1) \oplus (\bar{3}, 1, 1)(2) \oplus (1, 3, 2)(1) \oplus (1, 3, 1)(-2)
\end{equation}
\begin{multline}\label{eq10}
    78 = (8, 1, 1)(0) \oplus (1, 8 ,1)(0) \oplus (1, 1, 1)(0) \oplus \\ (1, 1, 2)(-3) \oplus (1, 1, 2)(3) \oplus (1, 1, 3)(0) \oplus (3, 3, 2)(-1) \\ \oplus (3, 3, 1)(2) \oplus (\bar3, \bar3, 2)(1) \oplus (\bar3, \bar3, 1)(-2)
\end{multline}
and, assembling the full chain $E_8 \rightarrow SU(3)_{spacetime,1} \otimes SU(3)_{genL} \otimes SU(3)_{c} \otimes SU(2)_{L} \otimes U(1)_{\gamma_1}$,
\begin{multline}\label{eq11}
    248 = (8, 1, 1, 1)(0) \oplus (1, 8, 1, 1)(0) \oplus (1, 1, 8 ,1)(0) \oplus \\
    (1, 1, 1, 1)(0) \oplus  (1, 1, 1, 2)(-3) \oplus (1, 1, 1, 2)(3) \oplus \\
    (1, 1, 1, 3)(0) \oplus (1, 3, 3, 2)(-1)  \oplus (1, 3, 3, 1)(2) \oplus \\
    (1, \bar3, \bar3, 2)(1) \oplus (1, \bar3, \bar3, 1)(-2) \oplus (3, \bar{3}, 3, 1)(0) \oplus \\
    \hspace{2mm}(3, 3, 1, 2)(1) \oplus (3, 3, 1, 1)(-2) \oplus (3, 1, \bar{3}, 2)(-1) \oplus\\
    (3, 1, \bar{3}, 1)(2) \oplus (\bar{3},3, \bar{3}, 1)( 0) \oplus
    (\bar{3}, \bar{3}, 1, 2)(-1) \oplus\\
    \hspace{-5mm}(\bar{3}, \bar{3}, 1, 1)(2) \oplus
    (\bar{3}, 1, 3, 2)(1) \oplus (\bar{3}, 1, 3, 1)(-2)
\end{multline}
The second $E_8$ branches identically, with $L \leftrightarrow R$ and $\gamma_1 \rightarrow \gamma_2$. The dimensions check: the three octets contribute $24$; the $SU(2)\otimes U(1)$ sector $(1+4+3)$ contributes $8$; the two conjugate $(3,\bar3,3,1)$-type sectors contribute $54$; and the three conjugate-paired families of mixed sectors contribute $3 \times 54$; in all, $24+8+54+162 = 248$.

One labelling convention must be declared before the comparisons of the next subsections, because the branching fixes the pairing of colour representation and abelian eigenvalue within each sector: we identify the physical quark colour with the $\bar{3}$ of the trinification colour slot (equivalently, we read the colour factor through its outer automorphism), so that the ledger sector $(\bar3,\bar3,2)(\gamma_1{=}1)$ carries the quark-doublet content $(3_c, 2)_{+1/6}$ and its conjugate carries $(\bar3_c,2)_{-1/6}$. This is a convention about names, with no physical content, but without it the colour and hypercharge entries of the ledger pair up with the opposite sign correlation to the standard-model one.

\subsection{The abelian charges: normalisation and origin}
\label{sec:hypercharge}

The integers in parentheses in Eqs.\ (\ref{eq6})--(\ref{eq11}) are eigenvalues of the trinification generator $U(1)_{\gamma}$ in a fixed integral normalisation. For the comparison with physical hypercharge assignments we use the eigenvalue dictionary
\begin{equation}
Y \;=\; \frac{\gamma_1}{2N}\,, \qquad N = \hbox{dimension of the $SU(3)_c$ representation of the sector},
\label{eq:dictionary}
\end{equation}
which converts, sector by sector, the integral eigenvalues into the standard hypercharge values ($\gamma_1 = 1$ on a colour triplet gives $Y = \tfrac16$, and so on). Equation (\ref{eq:dictionary}) is a \emph{dictionary between label conventions}, used only to compare branching labels with the physical roster; it is not the definition of a gauge generator, since a generator is a fixed Lie-algebra element and cannot be rescaled representation by representation. The physical origin and normalisation of the abelian charges is Clifford-algebraic and lies in the spinor lineage: electric charge is primary, $Q = N_o/3$ from the $Cl(6)$ number operator (Section~\ref{sec:cl6}), and hypercharge is the derived combination
\begin{equation}
Y \;=\; Q - T^3_L\,,
\label{eq:YQT}
\end{equation}
which reproduces every standard-model assignment, including $Y(\nu_R) = 0$, using no right-sector generator \cite{strongcp}. (Our electric-charge convention is $Q = Y + T^3_L$ with $T^3_L = \pm\tfrac12$ on doublets.) Whether the relative normalisation of $T^3_L$ and $Q$ in (\ref{eq:YQT}) is itself fixed by the construction, rather than adopted from the standard model, is open problem (O7).

\subsection{Reading the ledger: label assignments}
\label{sec:labels}

We now record how the physical roster is matched against the branching labels. One point must be made before any assignment is read: \emph{a branching determines representation content only}. It cannot determine spin or statistics, and it does not by itself distinguish a gauge field from a fermion from a scalar. The assignments below are therefore label correspondences --- the statement that a sector of (\ref{eq11}) carries the same $(SU(3)_{gen}, SU(3)_c, SU(2), U(1))$ content as a listed element of the physical roster --- and nothing more. The physical spin-statistics assignments reside in the spinorial and dynamical realisations (Sections~\ref{sec:fermions}, \ref{sec:dynamics}), not in the ledger.

With the dictionary (\ref{eq:dictionary}) understood, the visible-sector matches are as follows, writing $(a,b,c)(Y)$ for $(SU(3)_{genL}, SU(3)_c, SU(2)_L)$ content and physical hypercharge:
\begin{itemize}
\item[--] left-handed quark doublets $(u,d)_L$: $(\bar3,\bar3,2)(\tfrac16)$, from the $78$ [Eq.\ (\ref{eq10})]; their conjugates $(3,3,2)(-\tfrac16)$;
\item[--] left-handed lepton doublets $(\nu_e, e^-)_L$: $(\bar3,1,2)(-\tfrac12)$, from the $\overline{27}$ [Eq.\ (\ref{eq9})]; their conjugates $(3,1,2)(\tfrac12)$ from the $27$;
\item[--] gauge labels: $(1,8,1)(0)$ for the gluons, $(1,1,3)(0)$ for the $SU(2)_L$ triplet, $(1,1,1)(0)$ for the $U(1)_Y$ boson;
\item[--] a Higgs channel: the pair of colour-singlet weak-coset doublets of the trinification adjoint, $(1,1,2)(\gamma_1 = \pm 3)$. We note explicitly that the dictionary (\ref{eq:dictionary}) applied to these labels gives $\pm\tfrac32$, not the physical Higgs hypercharge $\pm\tfrac12$: the physical Higgs quantum numbers are fixed in the dynamical realisation (Section~\ref{sec:dynamics}), and this mismatch is one concrete illustration of the label--physics gap which the matching rule of Section~\ref{sec:ledger} bridges by fiat rather than by derivation.
\end{itemize}
The electric charges follow from $Q = Y + T^3_L$: $q_u = \tfrac16+\tfrac12 = \tfrac23$, $q_d = \tfrac16 - \tfrac12 = -\tfrac13$, $q_\nu = 0$, $q_e = -1$. Counting labels with their multiplicities and conjugates, and including the $SU(3)_{spacetime}$ multiplicity visible in (\ref{eq11}) (the quark labels are $SU(3)_{spacetime}$ singlets, the lepton labels triplets --- a triplet index with no counterpart on the spinorial roster, declared as exemption (a) of Section~\ref{sec:ledger}), the visible factor matches $36$ quark labels, $36$ lepton labels, $16$ boson labels, $8$ generator labels of $SU(3)_{spacetime,1}$ and $8$ of $SU(3)_{genL}$: $104$ labels in all.

The pre-gravitational factor is read mirror-wise: $SU(2)_R \otimes U(1)_{\gamma_2}$ in place of $SU(2)_L \otimes U(1)_{\gamma_1}$, with the $(1,8,1)(0)$ sector now the label of the \emph{same} colour octet seen in the second octonionic frame (Section~\ref{sec:onecolour}), and with the right-handed fermion tower $\nu_R$ (colour singlet), $d_R$, $u_R$ (colour triplets), $e^-_R$ (colour singlet). Its Higgs-channel doublet is the seed of the second Higgs doublet of the framework (Sections~\ref{sec:dynamics}, \ref{sec:outlook}). The mirror factor likewise matches $104$ labels. One clarification prevents a misreading: the $U(1)_{\gamma_2}$ eigenvalues carried by the mirror-factor fermion labels are ledger labels of the pre-gravitational lineage, not physical dark-electromagnetic charges of the visible fermions --- the visible fermions are dem-neutral (Section~\ref{sec:anomalies}) --- and the reconciliation of these mirror labels with the physical charge assignments is part of the identification problem (O1).

\subsection{The $208+288$ arithmetic, and the residual sector}
\label{sec:ledger}

The matching rule of the previous subsection is explicit, and we state it together with its two exemptions, which we display rather than hide: \emph{a sector of the branching is ``matched'' when its label content, under the stated dictionary and the colour convention of Section~\ref{sec:chain}, coincides with that of an element of the physical roster} --- the roster being the sixteen Weyl fermions per generation of Section~\ref{sec:fermions} with their three generations, the twelve gauge-boson labels per sector, the framework's two Higgs doublets (one seeded per sector, each compared together with its conjugate), and the coordinate-sector generators --- \emph{subject to two declared exemptions}: (a) the lepton labels carry an $SU(3)_{spacetime}$ triplet index which has no counterpart on the spinorial roster, whose states are $SU(3)_{spacetime}$ singlets, and is retained in the count as ledger multiplicity; (b) the Higgs-channel labels are matched by their $(SU(3)_{gen}, SU(3)_c, SU(2))$ content only, their dictionary hypercharge ($\pm\tfrac32$) differing from the physical value ($\pm\tfrac12$), which is fixed dynamically. Both exemptions are instances of the label--physics gap and belong to open problem (O3); for transparency, a rule without exemption (b) would match $100$ labels per factor, and a strict full-content rule without either exemption would match $64$, the lepton sectors failing wholesale on their spacetime index. Under the rule as stated, $104$ labels per factor are matched, $208$ in all, and the remaining $288$ are not. Per $E_8$, the unmatched labels form ten sectors, five plus their conjugates; the five independent ones are
\begin{equation}
(1,3,3,1)(2) \qquad (3, \bar{3}, 3, 1)(0) \qquad (\bar{3}, \bar{3}, 1, 1)(2) \qquad
(\bar{3}, 1, 3, 2)(1) \qquad (3, 1, \bar{3}, 1)(2)\,,
\label{eq:residual}
\end{equation}
$144$ labels per factor, $288$ in all, and $208+288=496$.

Three statements fix the status of this arithmetic, and we make them without qualification. (i) The matching rule is a \emph{defined comparison}, not a canonical representation-theoretic decomposition: there is no invariant of $E_8\times E_8$ which selects the $208$; the selection is made by comparison with an externally supplied roster. (ii) Consequently the $208$ is a count of matched \emph{labels}, not of physical degrees of freedom, and the $288$ is not a count of missing particles. (iii) The residual sector is physically silent in this framework: the $496$ is the dimension of an adjoint-lineage ledger, the physical matter lives in the spinor lineage, and an analysis of the fermion-bilinear content of the dynamics \cite{residual288} shows that the bifermionic seed, which lives in $27\otimes\overline{27} = 1\oplus 78\oplus 650$ of $E_6$, contains no channel that could source the sectors (\ref{eq:residual}): three of the five are excluded outright (no charge-sum channel, no fundamental generation triplet in the seed), and the remaining two sectors, which carry $SU(3)_{spacetime}$ charge, cannot be matter bilinears because the spinorial matter is an $SU(3)_{spacetime}$ singlet. What the bifermionic seed does supply, through its $78_L$ and $78_R$ channels, are the gauge currents and two composite electroweak Higgs doublets \cite{residual288, secondhiggs}. The \emph{visible-sector} matter and scalar content of the framework beyond the standard model is thereby fixed --- right-handed sterile neutrinos, one per generation, and a second Higgs doublet, with no superpartners, no fourth generation, and no light exotic coloured or charged states --- conditional on the bifermionic seed of \cite{residual288} exhausting the sources of physical states, which is part of open problems (O1) and (O8); the dark, dem-charged spectrum remains unfixed (Section~\ref{sec:anomalies}, open problem (O6)). This also reframes the familiar objection that $E_8\times E_8$ is ``too large'': on the present ontology the $496$ counts ledger labels, not particles. The force of the reframing is tied to the status of the partition rule (O3), and the ledger's irreducible role should be stated alongside it: the $\omega$-pairing of the two sectors, the mirror doubling, the generation-counting $SU(3)$, and the coordinate-sector $SU(3)$ are structures the spinor lineage alone does not supply, and they are what would be lost if the ledger were deleted.

\subsection{A single vector-like colour}
\label{sec:onecolour}

\begin{figure}[t]
    \centering
    \includegraphics[width=0.98\textwidth]{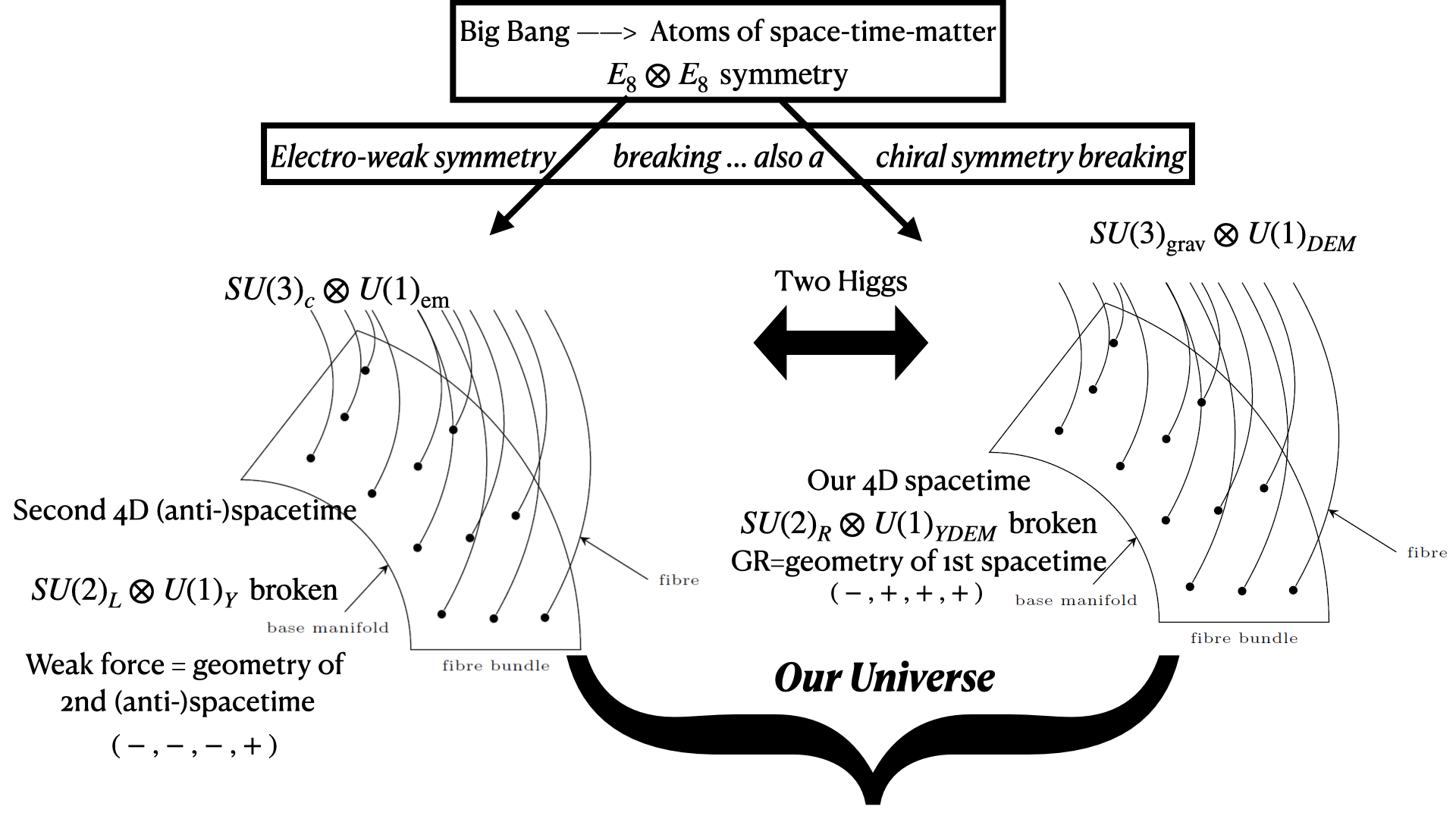}
    \caption{Schematic of the proposed emergence. The $E_8\times\omega E_8$ scaffolding organises two mirror sectors. Upon the (combined electroweak and left--right) symmetry breaking, one four-dimensional leaf carries the standard-model residue $SU(3)_c\times U(1)_{em}$, the other carries the pre-gravitational residue $U(1)_{dem}$ together with emergent general relativity; the weak interaction is the geometry of the flipped-signature leaf. $SU(3)_{grav}$ in the figure denotes the second frame-restriction of the single vector-like colour $SU(3)_c$ (Section~\ref{sec:onecolour}), not an independent gauge group. The two Higgs doublets couple the sectors.}
    \label{fig:unification}
\end{figure}

The branching of each factor contains a colour octet label, and a hasty reading would conclude that the framework has two colour groups. It has one. The left--right operation relating the two $E_6$ sectors is, in this framework, a spacetime (parity) operation --- it exchanges the two leaves of the six-dimensional arena of Section~\ref{sec:arena} --- and a spacetime operation cannot reach into the internal sector and double the colour living there \cite{strongcp}. Geometrically, colour is a single bundle over the common six-dimensional base, restricted to the two leaves; the two octet labels are the two frame-restrictions of one and the same octet. Consequently the physical colour group is one vector-like $SU(3)_c$: the right-handed quarks are triplets of the same $SU(3)_c$ as their left-handed partners, and both chiral electrons are colour singlets. (The second frame-restriction is the object labelled $SU(3)_{grav}$ in Figure~\ref{fig:unification}, which summarises the overall structure.) We flag the honest status of this statement: within the present framework it is a \emph{geometric postulate}, motivated by the parity reading of the left--right exchange and consistent with the anomaly analysis of \cite{strongcp}; a dynamical mechanism enforcing the diagonal identification (in the manner of a link field or a diagonal-breaking condensate in two-site models) has not been constructed, and is part of open problem (O2).

\section{Three generations, triality, and the exceptional Jordan algebra}
\label{sec:generations}

The branching of Section~\ref{sec:chain} supplies, in each factor, an $SU(3)_{gen}$ under which some ledger labels transform as triplets. The proposed account of physical triplication instead combines the triality of $Spin(8)$ with a Jordan frame in the Albert algebra. For an ordered frame, the three off-diagonal Peirce spaces $V_{12},V_{23},V_{31}$ are the three inequivalent eight-dimensional modules of its $Spin(8)$ stabiliser; the normaliser of the frame adds an $S_3$ which permutes them and induces the outer triality automorphisms. Identifying these modules with the three observed generations is a model postulate, not a consequence of triality alone \cite{Gresnigt, Gresnigt:2023ofp, Liam, Boyle, Todorov, BoyleF,fsc,mass}.

Two group-theoretic facts, frequently conflated in the literature, should be stated exactly. The automorphism group of $J_3(\mathbb{O})$ --- the transformations preserving the Jordan product --- is the compact exceptional group $F_4$, whose maximal subgroup $(SU(3)\otimes SU(3))/\mathbb{Z}_3$ splits, in the reading used here, into colour and generation factors \cite{Boyle, Todorov}. The complexified algebra $J_3(\mathbb{O})_{\mathbb{C}}$ has automorphism group $F_4(\mathbb{C})$; the group $E_6(\mathbb{C})$ is its \emph{reduced structure group}, the invariance group of the cubic norm (determinant), strictly larger than the automorphism group and preserving the determinant but not the Jordan product. It is in this determinant-preserving role that $E_6$ enters the fermionic sector of this programme \cite{fsc, mass}.

The physical output of this sector is spectral. The characteristic cubic of $J_3(\mathbb{O})$, evaluated on the appropriate charge sectors, yields the square-root mass ratios of the charged fermions: the trace split $1:2:3$ across the three charged sectors gives
\begin{equation}
\sqrt{m_e} : \sqrt{m_u} : \sqrt{m_d} \;=\; \tfrac13 : \tfrac23 : 1\,,
\label{eq:sqrtm}
\end{equation}
with a universal intra-family spread $\delta^2 = 3/8$, from which the charged-fermion mass ratios are obtained in closed form \cite{massratios}. Three qualifications, each established in \cite{strongcp}, bound the interpretation. First, $\sqrt{m}$ is not the charge of a gauged $U(1)$ acting on the visible fermions: on the visible fermion content, no anomaly-free abelian charge can carry the square-root-mass pattern, for any chirality structure or sign convention. It is a colour-blind spectral label of the exceptional Jordan mass operator. Second, the seating of the values $(0, \tfrac13, \tfrac23, 1)$ on the species $(\nu, e, u, d)$ is at present empirically motivated input; the one-third quantisation of $\sqrt{m}$ has no derivation yet, and the anomaly analysis constrains any future derivation to be spectral rather than gauge-theoretic. Third, the apparent electron--down interchange between the electric-charge assignments $(0,1,\tfrac23,\tfrac13)$ and the $\sqrt{m}$ assignments $(0,\tfrac13,\tfrac23,1)$ is a relation between two colour-blind functionals (a Dynkin-diagram swap in the flavour sector), not a permutation of states: no map can carry a one-dimensional colour representation onto a three-dimensional one, and the electron's colour representation is untouched. The CKM and PMNS matrices are, on this reading, the misalignment between the charge basis and the square-root-mass basis; first quantitative steps, including a computed quark CP phase and leptonic CP conservation with an inverted neutrino mass ordering, are reported in \cite{teliCKM, teliLepCP}.

\section{The six-dimensional arena}
\label{sec:arena}

\subsection{Octonionic coordinates and their symmetry}
\label{sec:coords}

In this framework the coordinate sector of each factor is an octonionic space: the eight directions $(1, e_1, \ldots, e_7)$ of $\mathbb{O}$ \cite{TPsir}. The automorphism group of $\mathbb{O}$ is $G_2$; fixing a unit imaginary octonion ($e_7$, the same choice that defines the ladder operators of Section~\ref{sec:cl6}) reduces it to the stabiliser subgroup $SU(3) \subset G_2$, and it is this $SU(3)$ which we call $SU(3)_{spacetime}$. Under it the octonionic coordinates transform as
\begin{equation}
\mathbb{O} \;\sim\; 1 \oplus 1 \oplus 3 \oplus \bar{3}
\label{eq:octdecomp}
\end{equation}
(the real direction and $e_7$ are singlets; the remaining six imaginary directions pair into a complex triplet and its conjugate). The $(8,1)$ sector of the branching (\ref{eq1}) is the set of \emph{generators} of this coordinate-sector symmetry; the coordinates themselves carry the reducible content (\ref{eq:octdecomp}). Inside $SU(3)_{spacetime}$ sits the $SU(2)$ whose $SO(3)$ quotient is the automorphism group of the quaternionic subalgebra $\langle 1, \hat i, \hat j, \hat k\rangle \subset \mathbb{O}$: the rotations of one quaternionic triple. It is at the quaternionic level, and through the split-complex unit, that spacetime signature appears, as we now describe. Colour, by contrast, requires the full octonion: the four imaginary directions beyond the quaternionic triple are the internal (colour) directions, so that the total count is $6+4 = 10$ --- but spacetime itself is six-dimensional, and the two extra dimensions beyond the familiar four are timelike, not spacelike.

\subsection{The split biquaternions and the $(3,3)$ quadratic form}
\label{sec:33}

The Clifford algebra $Cl(0,3)$, generated by three anticommuting imaginaries, is the algebra of \emph{split biquaternions} $\mathbb{H}\oplus\omega\mathbb{H}$: its volume element $\omega = e_1e_2e_3$ is central and squares to $+1$, and conjugation sends $\omega \to -\omega$. Between them the two blocks carry six distinct imaginary units, the quaternionic triple $(\hat i, \hat j, \hat k)$ and its split-dressed partner $(\omega\hat i, \omega\hat j, \omega\hat k)$. We coordinatise a six-dimensional space by
\begin{equation}
x_6 \;=\; t_1 \hat i + t_2 \hat j + t_3 \hat k + \omega\left( x_1 \hat i + x_2 \hat j + x_3 \hat k \right).
\label{eq:x6}
\end{equation}
With $\tilde{x}_6$ the split-biquaternionic conjugate (quaternion conjugation together with $\omega \to -\omega$), the symmetrised modulus is a scalar:
\begin{equation}
\tfrac{1}{2}\left\{ x_6 , \tilde{x}_6 \right\} \;=\; t_1^2 + t_2^2 + t_3^2 - x_1^2 - x_2^2 - x_3^2\,.
\label{eq:33form}
\end{equation}
This is a genuine quadratic form of signature $(3,3)$ on the six-dimensional real span of the imaginary units, and it is the origin of the Lorentzian structure in this framework: the minus signs are generated algebraically by the split unit, not inserted by hand. The elementary identity $e^{\omega\phi} = \cosh\phi + \omega\sinh\phi$ is the one-dimensional germ of the same fact: multiplication by a unit split-complex number is a boost.

The symmetry structure of $(\ref{eq:33form})$ must be stated with care, and less of it is algebraic than a first glance suggests; we state exactly how much. Conjugation $q \mapsto u q \bar u$ by a unit quaternion rotates the two triples \emph{simultaneously}: since $\omega$ is central, $u(\omega\hat e)\bar u = \omega\, u\hat e \bar u$, so the algebra automorphisms realise only the \emph{diagonal} $SO(3)$, rotating the $t$- and $x$-triples together, and this diagonal action is isometric. The independent rotations of the two triples --- which complete the maximal compact subgroup of the isometry group, $S(O(3)\times O(3))$, with identity component $SO(3)\times SO(3)$ --- are \emph{not} realised by any quaternionic or octonionic operation: no algebra automorphism rotates one triple without the other. Neither, a fortiori, are the boosts mixing the triples: no homomorphism can produce a non-compact simple group from compact ones. All of these belong to the frame group of the quadratic space defined by (\ref{eq:33form}): the group $SO(3,3)$, available exactly because the $\omega$-grading supplies the split quadratic form. The division of labour is thus sharper than one might hope, and we prefer to state it sharply: the algebra supplies the quadratic form and one diagonal rotation group; the frame group of that form supplies everything else. In particular, the two $SU(3)_{spacetime}$ factors of the branching do \emph{not} break to, and are not the origin of, $SO(3,3)$; their role is confined to the coordinate-sector symmetry described in Section~\ref{sec:coords}, of which the two $SO(3) \subset SO(3,3)$ rotation factors are the quaternionic restrictions. A Lorentzian world is obtained from compact ingredients plus one split sign --- and the split sign does all the non-compact work.

\subsection{Two four-dimensional leaves and their gauged dynamics}
\label{sec:leaves}

Inside the six-dimensional space $M_6$ with metric (\ref{eq:33form}) live two overlapping four-dimensional Lorentzian planes. Choosing unit imaginaries $t_L$ and $t_R$ in the undressed triple $\{\hat i,\hat j,\hat k\}$ --- the dressed partner $\omega t_R$ then supplying the odd direction of the second leaf ---
\begin{equation}
M_4^{(R)} := \omega\,{\rm span}\{\hat i,\hat j,\hat k\} \oplus {\rm span}\{t_L\} \quad (1,3), \qquad
M_4^{(L)} := {\rm span}\{\hat i,\hat j,\hat k\} \oplus {\rm span}\{\omega t_R\} \quad (3,1),
\label{eq:leaves}
\end{equation}
each borrowing its odd-signature direction from the opposite grading: the dressed triple is completed by the undressed $t_L$, and the undressed triple by the dressed $\omega t_R$. The two leaves overlap in the neutral $(1,1)$ plane ${\rm span}\{t_L, \omega t_R\}$: a two-dimensional bridge shared by the two four-dimensional worlds. Our universe is one leaf, curved by gravitation; the second, flipped-signature leaf carries the weak interaction as its geometry --- the weak force, on this reading, is a spacetime symmetry masquerading as an internal one, and its chirality is to be traced to the leaf assignment of the chiral spinor blocks (Section~\ref{sec:chirality}); demonstrating that this assignment reproduces the observed, exclusively left-handed weak coupling is part of open problem (O1). Related two-spacetime and $(3,3)$ structures have a substantial literature \cite{rf3patty, rf3everett, rf3cole, rf3teli, rf3kritov, rf3dartora, rf3podo, rf3batista, rf3boyling, rf3brody, Chester, rf3sparling, rf3mason, rf3lambek3, Pavsic1, Pavsic2, Pavsic3}.

Selecting a leaf must not destroy relativity, and the mechanism which ensures this also supplies the dynamics. Fix a negative-definite two-plane $N \subset M_6$ (the two discarded directions) and let $W = N^\perp$ be the surviving leaf. The stabiliser of $N$ in $SO(3,3)$ is $S(O(3,1)\times O(2))$, with identity component $SO(3,1)\times SO(2)$, and at the Lie-algebra level
\begin{equation}
\mathfrak{so}(3,3) \;=\; \underbrace{\mathfrak{so}(3,1)}_{6} \oplus \underbrace{\mathfrak{so}(2)}_{1} \oplus \underbrace{(W\wedge N)}_{8}\,,
\label{eq:sodecomp}
\end{equation}
so that promoting the normal \emph{two-plane} to a dynamical order parameter coupled to the six-dimensional spin connection can break $SO(3,3)\to SO(3,1)\times SO(2)$, with $15-7=8$ coset modes. If an ordered two-frame is used instead, it must carry a local right $SO(2)$ redundancy, or enter only through its plane projector; otherwise the normal $SO(2)$ is also broken and the Goldstone count is nine. One structural requirement should be made explicit here rather than left implicit: the rank-three bundles supplied by the $SU(2)\times U(1)$ reductions of the two coordinate-sector $SU(3)$s are associated bundles, and identifying their direct sum with the tangent bundle requires a soldering form $e:TM_6\xrightarrow{\sim}E_L\oplus E_R$. The $BF$ variables of \cite{Wesley} are a candidate supplier, but an explicit construction is part of open problem (O4). The further claim that constrained $BF$/Pleba\'nski sectors yield Einstein gravity on one leaf and weak isospin on the other is imported from \cite{Wesley} with its assumptions; the constraint choice, matter coupling and identification with the trinification $SU(2)$s are not derived here. Related gravi--weak ideas appear in \cite{Dehnen, Krasnov, Smolin, Marciano, Percacci, woit1, woit2, lqg}.

Two abelian remarks complete the picture. The only abelian factor the frame sector supplies is the $SO(2)$ of (\ref{eq:sodecomp}), which rotates the two discarded directions; it is not a hypercharge, and the analysis of \cite{Wesley} shows the two hypercharges $U(1)_Y$ and $U(1)_{Y_{dem}}$ cannot be extracted from the frame sector. They descend instead from the trinification chain of Section~\ref{sec:chain}, one per factor --- which is precisely why the scaffolding needs the internal $SU(3)^{\otimes 3}$ chain alongside the frame group. After the breaking, the visible chain terminates in $U(1)_{em}$ and the mirror chain in $U(1)_{dem}$: a dark-sector electromagnetism \cite{dark1, dark2, darkexp}, whose charged matter is dark (Section~\ref{sec:anomalies}), kinetically mixable with the visible photon, and whose astrophysical signatures are examined in \cite{finster, catalogue}.

\subsection{The gradient operator, and its exact mathematical status}
\label{sec:D6}

On the space (\ref{eq:x6}) the natural quaternionic gradient is
\begin{equation}
D_6 \;=\; \hat i\,\partial_{t_1} + \hat j\,\partial_{t_2} + \hat k\,\partial_{t_3} + \omega\left( \hat i\,\partial_{x_1} + \hat j\,\partial_{x_2} + \hat k\,\partial_{x_3} \right) \;\equiv\; D_L + \omega D_R\,,
\label{eq:D6}
\end{equation}
a six-dimensional generalisation of Hamilton's operator $\nabla = \hat i\,\partial_1 + \hat j\,\partial_2 + \hat k\,\partial_3$, whose square is minus the Laplacian --- the observation which led Atiyah to remark that the Dirac operator was first found by Hamilton \cite{rf3Atiyah}. Symmetrised against its conjugate $\tilde D_6 = -D_L + \omega D_R$, the operator returns the wave operator of the $(3,3)$ metric:
\begin{equation}
\tfrac{1}{2}\left\{ D_6 , \tilde{D}_6 \right\} \;=\; \partial_{t_1}^2 + \partial_{t_2}^2 + \partial_{t_3}^2 - \partial_{x_1}^2 - \partial_{x_2}^2 - \partial_{x_3}^2 \;=\; \Box_{3,3}\,.
\label{eq:D6sq}
\end{equation}
Equation (\ref{eq:D6sq}) is an exact operator identity, the differential-operator counterpart of (\ref{eq:33form}).

The mathematical status of $D_6$ must be stated exactly, and we do so here rather than leave it implicit. \emph{The six coefficients of $D_6$ do not form a Clifford generating set for $Cl(3,3)$.} Within each triple the units anticommute, but each unit \emph{commutes} with its own dressed partner, $\{\hat i, \omega\hat i\} = -2\omega \neq 0$, since $\omega$ is central; and on dimensional grounds no such set could close, since $Cl(3,3) \cong M_8(\mathbb{R})$ is sixty-four-dimensional while the split-biquaternion algebra is eight-dimensional. Equation (\ref{eq:D6sq}) is therefore a \emph{split-biquaternionic factorisation of the wave operator through conjugation and symmetrisation}, and not the usual Clifford factorisation by six mutually anticommuting gamma matrices; correspondingly, the ordered products differ from the symmetrised one by an operator-valued remnant,
\begin{equation}
D_6 \tilde D_6 \;=\; \Box_{3,3} + \omega\,[D_L, D_R]\,, \qquad
\tilde D_6 D_6 \;=\; \Box_{3,3} - \omega\,[D_L, D_R]\,,
\label{eq:remnant}
\end{equation}
with $[D_L, D_R] = \sum_{a\neq b} [\hat e_a, \hat e_b]\,\partial_{t_a}\partial_{x_b} \neq 0$. Because $D_6$ is not Hermitian, its first-order equation is properly a conjugate pair,
\begin{equation}
i\hbar\, D_6\,\psi = Qc\,\psi\,, \qquad i\hbar\, \tilde{D}_6\,\tilde{\psi} = Qc\,\tilde{\psi}\,,
\label{eq:pair}
\end{equation}
in which the two members are conjugate partners --- we note that the anti-automorphism reverses products, so the pair is posed jointly, the second member being the conjugate-partner equation with the left-acting $\tilde D_6$, not an operator-by-operator image of the first; the source $Q$ is a general charge of the unbroken gravi-weak sector, from which electric charge and mass are both emergent after the breaking. Mass-shell propagation must then be stated exactly. Since
\begin{equation}
[D_L, D_R] \;=\; \sum_{a<b} [\hat e_a, \hat e_b]\left(\partial_{t_a}\partial_{x_b} - \partial_{t_b}\partial_{x_a}\right),
\label{eq:kernel}
\end{equation}
and the three commutators $[\hat e_a,\hat e_b] = 2\hat e_c$ are linearly independent, the kernel of the remnant consists exactly of the fields with symmetric mixed Hessian, $\partial_{t_a}\partial_{x_b}\psi = \partial_{t_b}\partial_{x_a}\psi$ for all pairs: mode by mode, plane waves whose temporal and spatial wave-vectors are parallel. On this kernel $[D_6, \tilde D_6]\psi = 2\omega[D_L,D_R]\psi = 0$, so the two operators may be diagonalised jointly, and a field satisfying \emph{both} members of the pair simultaneously, $D_6\psi = \tilde D_6 \psi = (Qc/i\hbar)\psi$, obeys the Klein--Gordon equation $\Box_{3,3}\psi = -(Qc/\hbar)^2\psi$ by (\ref{eq:D6sq}); we emphasise that kernel membership together with the first member alone does not suffice, the joint condition being part of the definition of a physical solution. Every field varying only along a \emph{matched} pair --- the two-plane ${\rm span}\{\hat e, \omega\hat e\}$ of a single unit $\hat e$, which is the bridge of Section~\ref{sec:leaves} with $t_R$ chosen parallel to $t_L$ --- lies in the kernel; indeed on a matched pair the factorisation is exact even as an \emph{ordered} product, $(\hat i\,\partial_t + \omega\hat i\,\partial_x)(-\hat i\,\partial_t + \omega\hat i\,\partial_x) = \partial_t^2 - \partial_x^2$, with no symmetrisation needed. (A bridge with $t_R$ not parallel to $t_L$ is not matched, and its fields are not in the kernel.) But a general field on $M_6$ --- and, we stress, even a general field on a four-dimensional leaf of (\ref{eq:leaves}) --- is \emph{not} in the kernel: the naive restriction of $D_6$ to a $(1,3)$ leaf retains a remnant in its ordered products. The operator $D_6$ therefore does not by itself supply the leaf Dirac theory; the familiar four-dimensional Dirac operator belongs to the doubled (gamma-matrix) representation, and the honest statement for the unreduced theory is that a genuine six-dimensional Dirac operator requires the Clifford algebra $Cl(3,3) \cong M_8(\mathbb{R})$ represented on an eight-real-dimensional spinor module --- for which the split-biquaternion algebra itself, of real dimension eight, is the natural candidate carrier, with the gamma matrices realised as elements of its endomorphism algebra rather than of the algebra itself. That the endomorphism (multiplication) algebras of the division algebras naturally furnish Clifford structures is a known and currently active theme --- the full multiplication algebra of the quaternions is $Cl(3,1)$ \cite{FureyNested}. The required construction is carried out in the next subsection, on a different and smaller carrier: the split quaternions. $D_6$ itself retains the status of a symmetrised square root of $\Box_{3,3}$ --- exact as (\ref{eq:D6sq}), exact as a first-order factorisation on the $(1,1)$ bridge, and no more is claimed for it --- while its exact relation to the spinor-level operator of Section~\ref{sec:cl33} is part of open problem (O4). It is worth remarking that the kernel restriction just derived --- alignment of temporal and spatial gradients --- is of the same character as the data restrictions under which ultrahyperbolic equations admit a well-posed problem \cite{CraigWeinstein}; we do not pursue the connection here.

One further caution belongs here rather than in fine print. A wave equation on a $(3,3)$ background is ultrahyperbolic, and ultrahyperbolic equations do not possess the unconstrained Cauchy problem of a one-time theory: well-posedness holds only for data restricted in appropriate ways \cite{CraigWeinstein}. In the present framework the physical dynamics is posed on the emergent leaves, each of Lorentzian signature, and the six-dimensional arena is pre-geometric --- the evolution parameter of the fundamental dynamics is Connes time, external to the six coordinates (Section~\ref{sec:dynamics}) --- so no ultrahyperbolic initial-value problem is invoked for observed physics. But a complete treatment of the unbroken phase must confront this structure honestly; it is folded into open problem (O5).

\subsection{The exact $Cl(3,3)$ Dirac operator: split quaternions and the Jordan algebra $J_2(\mathbb{H}_s)$}
\label{sec:cl33}

The preceding subsection ended with a deficit: $D_6$ is a symmetrised square root of $\Box_{3,3}$, not a Clifford--Dirac operator, and its naive leaf restrictions fail. We now construct the operator that closes this deficit exactly. The carrier is not a further enlargement of the split biquaternions but a different algebra one rung over: the \emph{split quaternions} $\mathbb{H}_s$, spanned by $1, \mathbf{i}, \mathbf{j}, \mathbf{k}$ with
\begin{equation}
\mathbf{i}^2 = -1, \qquad \mathbf{j}^2 = \mathbf{k}^2 = +1, \qquad \mathbf{i}\mathbf{j} = \mathbf{k} = -\mathbf{j}\mathbf{i},
\label{eq:Hs}
\end{equation}
all three imaginary units mutually anticommuting. $\mathbb{H}_s$ is isomorphic to $M_2(\mathbb{R})$, i.e.\ to the Clifford algebra $Cl(1,1)$; it is a split composition algebra with zero divisors, not a division algebra, and the graded triple product $Cl(1,1)\,\widehat\otimes\,Cl(1,1)\,\widehat\otimes\,Cl(1,1) \cong Cl(3,3)$ already signals its role. For $q = a + b\mathbf{i} + c\mathbf{j} + d\mathbf{k}$ the conjugate is $\bar q = a - b\mathbf{i} - c\mathbf{j} - d\mathbf{k}$ and the norm $N(q) = q\bar q = a^2 + b^2 - c^2 - d^2$ has signature $(2,2)$.

Form the Jordan algebra $J_2(\mathbb{H}_s)$ of $2\times 2$ Hermitian matrices over $\mathbb{H}_s$ (real diagonal, conjugate off-diagonal entries, symmetrised product). Writing
\begin{equation}
X = \begin{pmatrix} t_1 + x_1 & q \\ \bar q & t_1 - x_1 \end{pmatrix}, \qquad
q = x_2 + \mathbf{i}\,x_3 + \mathbf{j}\,t_2 + \mathbf{k}\,t_3\,,
\label{eq:J2X}
\end{equation}
one finds
\begin{equation}
\det X \;=\; (t_1 + x_1)(t_1 - x_1) - N(q) \;=\; t_1^2 + t_2^2 + t_3^2 - x_1^2 - x_2^2 - x_3^2\,,
\label{eq:J2det}
\end{equation}
so that, as a quadratic space,
\begin{equation}
J_2(\mathbb{H}_s) \;\cong\; \mathbb{R}^{3,3}.
\label{eq:J2R33}
\end{equation}
This is the split-signature member of the classical family of identifications $J_2(\mathbb{C}) \cong \mathbb{R}^{1,3}$, $J_2(\mathbb{H}) \cong \mathbb{R}^{1,5}$, $J_2(\mathbb{O}) \cong \mathbb{R}^{1,9}$ that underlies the division-algebra description of Minkowski spacetimes \cite{KugoTownsend, Sudbery1984, Baez}, and an instance of the split-composition-algebra sector of the $2\times 2$ magic square \cite{BartonSudbery, DHK}; the isomorphisms $Spin(3,3) \cong SL(4,\mathbb{R})$ and $Cl(3,3)\cong M_8(\mathbb{R})$ invoked below are classical \cite{Harvey}.

Choose an algebra isomorphism $\rho:\mathbb H_s\to M_2(\mathbb R)$. Applying $\rho$ entrywise gives a real $4\times4$ map, denoted $\rho_2(X)$, between the two Weyl modules for every $X\in J_2(\mathbb H_s)$. On this space define the split-quaternionic matrix gradient and its adjugate,
\begin{equation}
\nabla_{3,3} = \begin{pmatrix} \partial_{t_1} + \partial_{x_1} & D_q \\ \bar D_q & \partial_{t_1} - \partial_{x_1} \end{pmatrix}, \qquad
\widetilde\nabla_{3,3} = \begin{pmatrix} \partial_{t_1} - \partial_{x_1} & -D_q \\ -\bar D_q & \partial_{t_1} + \partial_{x_1} \end{pmatrix},
\label{eq:nabla33}
\end{equation}
with $D_q = \partial_{x_2} + \mathbf{i}\,\partial_{x_3} + \mathbf{j}\,\partial_{t_2} + \mathbf{k}\,\partial_{t_3}$ and $\bar D_q$ its conjugate. Because the three imaginary units of $\mathbb{H}_s$ mutually anticommute, $D_q \bar D_q = \bar D_q D_q = \partial_{x_2}^2 + \partial_{x_3}^2 - \partial_{t_2}^2 - \partial_{t_3}^2$ with no cross terms, and hence, by the $2\times 2$ adjugate identity,
\begin{equation}
\nabla_{3,3}\,\widetilde\nabla_{3,3} \;=\; \widetilde\nabla_{3,3}\,\nabla_{3,3} \;=\; \Box_{3,3}\,\mathbb{I}\,,
\label{eq:nablaexact}
\end{equation}
\emph{exactly, in both orderings, with no operator-valued remnant and no symmetrisation} --- the property that $D_6$ could deliver only on the kernel of its remnant. Here the displayed $2\times2$ split-quaternionic matrices are understood through $\rho$ as real $4\times4$ operators, and $\psi_\pm\in S_\pm\cong\mathbb R^4$. Treating $\psi_\pm$ instead as unrestricted columns in $\mathbb H_s^2$ would double the real module and yield a reducible representation. The corresponding Dirac system is the chiral pair
\begin{equation}
i\hbar\,\nabla_{3,3}\,\psi_- = Qc\,\psi_+\,, \qquad i\hbar\,\widetilde\nabla_{3,3}\,\psi_+ = Qc\,\psi_-\,,
\label{eq:chiralpair}
\end{equation}
and eliminating either component gives $-\hbar^2\Box_{3,3}\psi_\pm=Q^2c^2\psi_\pm$. Doubling the represented vector map,
\[
\Gamma(X)=\begin{pmatrix}0&\rho_2(X)\\ \rho_2(\widetilde X)&0\end{pmatrix},
\]
gives an $8\times8$ real matrix satisfying $\Gamma(X)^2=\det(X)\mathbb I$. By polarisation this is the Clifford relation, and the generated algebra is $Cl(3,3)\cong M_8(\mathbb R)$. Its even part is $M_4(\mathbb R)\oplus M_4(\mathbb R)$, with Weyl modules $S_\pm\cong\mathbb R^4$, and $Spin_0(3,3)\cong SL(4,\mathbb R)$. The two Weyl modules are not the two spacetime leaves; correlating chirality with leaf support is dynamical and remains part of (O1).

The decisive property of $\nabla_{3,3}$ for this framework --- and one we have verified by direct computation --- is its behaviour under restriction to the leaves of Section~\ref{sec:leaves}. Setting $\partial_{t_2} = \partial_{t_3} = 0$ (fields on the $(1,3)$ leaf) reduces $\nabla_{3,3}$ to the standard two-component Weyl operator of four-dimensional Minkowski spacetime, with ordered product $\partial_{t_1}^2 - \partial_{x_1}^2 - \partial_{x_2}^2 - \partial_{x_3}^2$ exactly; setting $\partial_{x_2} = \partial_{x_3} = 0$ reduces it to the flipped-signature $(3,1)$ operator, with ordered product $\partial_{t_1}^2 + \partial_{t_2}^2 + \partial_{t_3}^2 - \partial_{x_1}^2$; and the bridge restriction gives the $(1,1)$ operator. One operator thus supplies the six-dimensional wave operator, the Weyl operators of both leaves, and the bridge operator, all as exact identities --- precisely the structure the two-leaf gravi-weak picture requires, and precisely what the naive restrictions of $D_6$ failed to provide.

Three remarks fix the status and the attribution of this construction. \emph{First, attribution.} The $2\times 2$ Hermitian-matrix model with $SL(2,\mathbb{H}) \cong Spin(5,1)$ is standard for the division quaternions, and the two-component quaternionic spinor formalism for six-dimensional \emph{Lorentzian} spacetimes $\mathbb{R}^{1,5}$ (signature written $(5,1)$ in their convention) has been developed in detail by Ven\^ancio and Batista \cite{rf3batista}; the present construction is its split-signature analogue, obtained by the replacement $\mathbb{H} \to \mathbb{H}_s$. We have not found the $(3,3)$ split-quaternionic case written out in the literature --- earlier six-dimensional $(3,3)$ Dirac equations \cite{rf3patty, rf3boyling, rf3dartora} employ conventional complex gamma matrices --- but every ingredient is classical or an immediate corollary of published results \cite{Sudbery1984, BartonSudbery, DHK, Harvey}, and we present the construction in that spirit. \emph{Second, relation to $D_6$.} The two operators live in different layers: $D_6$ acts on the coordinate algebra --- it belongs to the construction of the base, whose $\omega$-grading supplies the quadratic form (\ref{eq:33form}) and the mirror pairing --- while $\nabla_{3,3}$ acts on the spinor modules of the Clifford algebra of that quadratic form. The two layers are dimensionally matched: the split-biquaternion algebra has real dimension eight, equal to $\dim_{\mathbb{R}}(S_+ \oplus S_-)$, so the split-biquaternion-valued field $\psi$ of (\ref{eq:pair}) can carry the $Cl(3,3)$ Dirac-spinor representation; constructing the explicit intertwiner is the sharpened residue of open problem (O4). \emph{Third, the Jordan-algebraic reading.} $J_2(\mathbb{H}_s)$ is a rank-two Jordan algebra whose norm form is quadratic, and its appearance here effects a clean division of labour with the cubic Jordan algebra of Section~\ref{sec:generations}: $J_2(\mathbb{H}_s)$ supplies the $(3,3)$ spacetime base and its Clifford principal symbol, while $J_3(\mathbb{O}_{\mathbb{C}})$ supplies the internal, generation and mass structure. The two are related inside the Albert algebra: with respect to a primitive idempotent, $J_3(\mathbb{O})$ has Peirce decomposition $27 = 1 + 16 + 10$, whose ten-dimensional block is $J_2(\mathbb{O}) \cong \mathbb{R}^{1,9}$ --- the rank-two corner of the rank-three algebra. One real-form subtlety must be stated: $J_2(\mathbb{H}_s)$, being indefinite, does not embed in the formally real (compact) $J_3(\mathbb{O})$; it embeds in the split Albert algebra $J_3(\mathbb{O}_s)$, and both real forms sit inside the complexification $J_3(\mathbb{O}_{\mathbb{C}})$ used in Section~\ref{sec:generations}. A complete treatment must therefore state which real structure selects the split $(3,3)$ spacetime sector and which selects the Hermitian mass sector; we flag this under (O4). A natural composite operator, schematically $D = \slashed{\partial}_{3,3} \otimes \mathbb{I} + \Gamma_7 \otimes D_{\rm int}$ with $D_{\rm int}$ built from the $J_3(\mathbb{O}_{\mathbb{C}})$ data, would act identically on the three generations through its spacetime part and distinguish them through its internal part; making this product structure precise --- in particular, exhibiting the commuting actions of $Cl(3,3)$ and the internal $Cl(6)$ on one module, which must not simply be identified --- belongs to open problems (O1) and (O4).

\paragraph{Real-form bookkeeping.}
The real-form issue is not cosmetic. One has $\mathrm{Str}_0J_3(\mathbb O)=E_{6(-26)}$ and $\mathrm{Str}_0J_3(\mathbb O_s)=E_{6(6)}$, whereas the compact $E_6$ in the compact branching $E_8\supset SU(3)\times E_6$ acts on the complex $\mathbf{27}$ while preserving an additional Hermitian form. We therefore use the complex magic-star decomposition as the common algebraic statement. A single compatible choice of real structure producing both the split spacetime slice and the desired internal Hermitian spectrum has not been constructed.

\section{The dynamical setting}
\label{sec:dynamics}

The scaffolding described so far is kinematic. The dynamics of the programme is a generalised trace dynamics \cite{Adler_1994, Adler, TPsir, review, review2}: the fundamental degrees of freedom are matrices (``aikyons'') valued in the split-bioctonionic extension of the division algebras, evolving in Connes time $\tau$ --- the modular time of the underlying operator algebra \cite{Connes_1996} --- with a trace Lagrangian bilinear in a matrix-valued Dirac-type operator. Bosonic (even-grade) matrix components carry the gauge and geometric content; fermionic (odd-grade, Grassmann) components carry the matter; the theory is pre-quantum and pre-spacetime, and both quantum field theory on the emergent leaves and classical general relativity are to be recovered in appropriate limits, the quantum-to-classical transition being driven by the entanglement of many aikyons and completed by spontaneous localisation.

Within this setting the following have been carried out in published work, and we cite them with their own qualifications. The symmetry breaking of the $SO(3,3)$ $BF$ theory, with its two Pleba\'nski sectors delivering Einstein gravity on one leaf and weak isospin on the other, is constructed in \cite{Wesley}. The recovery of the Einstein--Hilbert, Yang--Mills, Dirac and Higgs sectors from the generalised trace-dynamics Lagrangian, organised by the spectral-action principle \cite{3cc2, Connes_1996}, is presented in \cite{universeGTD}; that work describes itself as a conditional structural framework --- the heat-kernel coefficients are computed under stated assumptions on the spectral triple, and the construction is not a complete first-principles derivation --- and we import it here with exactly that status. The derivation of the charged-fermion mass ratios from the exceptional Jordan algebra is given in \cite{massratios}; the gauge-coupling analysis, including the equal-trace/two-channel-locking construction giving $\sin^2\theta_{W,\star} = 3/13 = 0.2308$ and a low-energy $\alpha_\gamma^{-1} \simeq 137.04$, is given in \cite{gaugecouplings}, whose own statement of its hypotheses (colour matching, exponential character, two-channel locking) applies verbatim; and the strong-CP analysis, with the anomaly results used in Sections~\ref{sec:generations} and \ref{sec:anomalies}, is given in \cite{strongcp}.

One structural statement about the dynamics belongs in this paper because it concerns the scaffolding directly. The trace-dynamics action is built from matrices whose components are organised by the label content of Section~\ref{sec:branching}; but an explicit realisation of $E_8\times E_8$ as a symmetry of that action --- a specified module on which the group acts, an invariant bilinear form, and the transformation law of the fermionic sector --- has not been constructed. At present, $E_8\times E_8$ functions as the organising symmetry of the label content, exact at the level of the ledger and conjectural at the level of the action. We regard the construction of the invariant action (or the demonstration that none exists, which would falsify the scaffolding hypothesis) as the central formal problem of the programme: open problem (O1).

\section{Anomaly structure}
\label{sec:anomalies}

The anomaly accounting separates into three layers, and the three have different logical characters: the first is standard and complete, the second is representation-theoretic and complete, the third is conditional and partly open.

\emph{Layer 1: the physical chiral sector, four-dimensional.} The fermions of one generation are the sixteen Weyl states of Section~\ref{sec:fermions}: the content of the $16$ of $Spin(10)$, i.e.\ one standard-model generation with its right-handed neutrino. For this content, the vanishing of all cubic gauge anomalies $[SU(3)_c]^3$, $[SU(2)_L]^2 U(1)_Y$, $[SU(3)_c]^2 U(1)_Y$, $[U(1)_Y]^3$, of the mixed gauge-gravitational anomaly $[{\rm grav}]^2 U(1)_Y$, and of the global Witten $SU(2)$ anomaly (four doublets per generation, an even number) is standard \cite{Witten1982, AGW}. Within this framework the hypercharges are not adjustable once the normalisation of $T^3_L$ relative to $Q$ is fixed: $Q = N_o/3$ is derived, $Y = Q - T^3_L$ then lands on the anomaly-free assignment (Section~\ref{sec:hypercharge}), and the faithfulness of that relative normalisation is precisely open problem (O7). Subject to (O7), anomaly cancellation functions as a consistency check which the spinorial realisation passes, generation by generation.

\emph{Layer 2: the ledger.} The adjoint of $E_8$ is a real (self-conjugate) representation, and in the branching (\ref{eq11}) every complex sector appears together with its conjugate. Consequently the total cubic anomaly of the full $496$ vanishes \emph{pairwise}: each sector's contribution is cancelled by its conjugate's. (The cancellation is between conjugate pairs, not within individual sectors: a single sector such as $(1,3,3,1)(2)$ has a non-vanishing cubic anomaly by itself.) This pairwise structure is the anomaly-theoretic face of the Distler--Garibaldi theorem \cite{DG}: fermionic matter placed in the adjoint lineage of $E_8$ is necessarily vector-like. It is exactly why the ledger cannot house the chiral matter, and why the framework houses the matter in the spinor lineage instead (Section~\ref{sec:ontology}). No ten-dimensional Green--Schwarz condition applies, because there is no ten-dimensional chiral field theory here: the arena is the six-dimensional $(3,3)$ structure of Section~\ref{sec:arena}, and observed physics lives on its four-dimensional leaves.

\emph{Layer 3: the abelian dark sector, and what remains conditional.} The one abelian gauge factor beyond the standard model is $U(1)_{dem}$. On the \emph{visible} fermion content, it is proven in \cite{strongcp} that no anomaly-free abelian charge can carry the square-root-mass pattern, for any chirality structure or sign convention; the visible fermions are therefore dem-neutral, coupling at most through kinetic mixing, and the dem-charged matter is dark. Two statements must then be kept conditional, and we so mark them. (i) Anomaly freedom of the full $U(1)_{dem}$ sector requires the dark-sector chiral spectrum and charges, which the framework constrains but has not yet fixed; until that spectrum is specified, dem anomaly freedom is a requirement on model building, not a result. (ii) The useful classification statement --- that on three visible generations every anomaly-free abelian charge lies in ${\rm span}\{Y, B-L\}$ --- holds under the assumptions of generation universality and compatibility with the Yukawa sector; charges such as $L_\mu - L_\tau$ evade it when universality is dropped. Both conditions are part of open problem (O6). Finally, if the six-dimensional arena carries chiral dynamics above the breaking scale, the six-dimensional anomaly structure (box anomalies: gauge, gravitational, and mixed) must be analysed in its own right; this analysis has not been performed and is also folded into (O6).

\section{Open problems}
\label{sec:open}

The framework is a programme with sharply posed problems, not a completed unification, and we prefer to state the problems plainly. In rough order of logical priority:

\begin{itemize}
\item[\textbf{(O1)}] \emph{The invariant action, and the identification map.} Construct an explicit realisation of $E_8\times E_8$ on the physical state space --- the spinor module of Section~\ref{sec:fermions} together with the geometric sector --- with an invariant bilinear form and an invariant coupling; or prove that no such realisation exists. At present the group acts on the ledger, not provably on the dynamics (Section~\ref{sec:dynamics}). The same problem contains the identification map of Sections~\ref{sec:intro} and \ref{sec:leaves}: realising the trinification $SU(2)$s as the chiral halves of the frame sector, reconciling the internal chiral-gauge characterisation of $SU(2)_R$ \cite{l-r} with its gravitational realisation, relating the mirror-factor ledger charges to the physical (dem-neutral) assignments, and demonstrating that the leaf assignment reproduces the observed left-handed weak coupling.
\item[\textbf{(O2)}] \emph{The diagonal colour.} Supply the dynamical mechanism (link field, condensate, or constraint) enforcing the identification of the two colour-octet frame-restrictions as one vector-like $SU(3)_c$ (Section~\ref{sec:onecolour}), which is at present a geometric postulate.
\item[\textbf{(O3)}] \emph{The partition rule.} The matched/residual split of the ledger ($208+288$) rests on a defined comparison with the physical roster (Section~\ref{sec:ledger}); likewise the map between the trinification labels and the $Cl(6)$ three-generation module is imported, not derived. Derive both, or establish that they are conventions without physical weight.
\item[\textbf{(O4)}] \emph{Completing the $Cl(3,3)$ Dirac layer.} The exact six-dimensional Dirac operator is constructed in Section~\ref{sec:cl33}; what remains is its integration. Exhibit the explicit intertwiner between the split-biquaternion-valued field of (\ref{eq:pair}) and the chiral modules $S_+ \oplus S_-$ (the real dimensions match); state which real structures of $J_3(\mathbb{O}_{\mathbb{C}})$ select the split spacetime sector and the Hermitian mass sector; extend $\nabla_{3,3}$ to the curved, soldered setting of Section~\ref{sec:leaves} (the multiplication-algebra route \cite{FureyNested} remains a natural alternative track); and embed the resulting operator in the trace-dynamics Lagrangian of Section~\ref{sec:dynamics}.
\item[\textbf{(O5)}] \emph{Consistency with three times.} Give a complete account of the unbroken $(3,3)$ phase consistent with the ultrahyperbolic character of its wave operators \cite{CraigWeinstein}: identify the constraint structure, or the mechanism (leaf localisation, Connes-time evolution) by which the physical theory never poses an ill-posed Cauchy problem; and demonstrate the absence of the standard multi-time pathologies --- ghost (negative-norm) excitations, tachyonic instabilities, and unitarity violations associated with the two additional timelike directions --- both in the unbroken phase and on the flipped-signature leaf, whose three timelike directions keep this question live even after leaf selection.
\item[\textbf{(O6)}] \emph{Anomalies beyond layer 1.} Specify the dark-sector chiral spectrum and verify $U(1)_{dem}$ anomaly freedom; carry out the six-dimensional (box) anomaly analysis of the arena; and state the classification of allowed abelian charges with its universality assumptions made explicit (Section~\ref{sec:anomalies}).
\item[\textbf{(O7)}] \emph{Normalisation faithfulness.} Establish that the relative normalisation of $T^3_L$ and $Q$ in $Y = Q - T^3_L$ is fixed by the construction rather than adopted (Section~\ref{sec:hypercharge}).
\item[\textbf{(O8)}] \emph{Scales and the Higgs sector.} The condensation of the electroweak-doublet channels of the bifermionic Lagrangian into the two physical Higgs doublets is argued, not established by a bound-state calculation; the doublet masses and mixing are underived; and the hierarchy $v/M_{Pl} \sim 10^{-17}$ is relocated (into the gravi-weak breaking), not solved \cite{residual288, secondhiggs}.
\item[\textbf{(O9)}] \emph{Flavour.} Derive the one-third quantisation of $\sqrt{m}$ and the seating of $(0,\tfrac13,\tfrac23,1)$ on $(\nu, e, u, d)$ (Section~\ref{sec:generations}); complete the CKM/PMNS derivation begun in \cite{teliCKM, teliLepCP}; establish the loop-level stability of the strong-CP result \cite{strongcp}.
\end{itemize}

\section{Experimental outlook}
\label{sec:outlook}

The framework makes no positive prediction of a new particle with a derived mass at collider energies, and we do not claim one. Its experimental content is nonetheless substantial, and is catalogued, graded by logical strength, in a falsification-oriented registry \cite{catalogue}; we summarise the principal entries.

At colliders, the one accessible new state is the second Higgs doublet (Sections~\ref{sec:branching}, \ref{sec:dynamics}). Its compositeness is undetectable --- form-factor signatures are suppressed by $v^2/M_{Pl}^2 \sim 10^{-34}$, so both scalars behave as effectively fundamental --- but the doublet itself is detectable in principle: the physical spectrum is a quartet (a CP-even scalar $H$, a CP-odd scalar $A$, and a charged pair $H^\pm$), described at low energy by a CP-conserving two-Higgs-doublet model close to the alignment limit with a Type-I-like Yukawa assignment \cite{secondhiggs}. The quartet mass is not determined by the theory; the electroweak-matching structure motivates, without forcing, the window $10^2$--$10^3$ GeV. The most robust discovery channel is electroweak pair production ($pp \to H^+H^-,\ H^\pm A,\ H^\pm H,\ HA$), whose rate is fixed by gauge quantum numbers alone and survives the alignment limit; indirect reach comes through Higgs-coupling precision at the HL-LHC and at an $e^+e^-$ Higgs factory. Beyond this, the framework predicts characteristic \emph{absences}: no superpartners, no fourth generation, no light exotic coloured or charged states, no right-handed $W$ boson, and no axion. Most of these coincide with the null results that currently favour the plain standard model and so do not by themselves discriminate the framework; the absence of $W_R$ signals, however, is a consequence of the gravi-weak identification (the right-sector $SU(2)$ realised gravitationally rather than internally) and shares that identification's conjectural status (Sections~\ref{sec:intro}, \ref{sec:leaves}) rather than being an independent choice \cite{strongcp}; and the absence of light exotics discriminates between the programme's internal scale branches \cite{residual288}.

The sharper near-term tests lie outside colliders: the parameter-free flavour relations descending from (\ref{eq:sqrtm}), whose confrontation with data requires a common-scale evaluation of running masses \cite{massratios}; the gauge-sector relations $\sin^2\theta_{W,\star} = 3/13$ and $\alpha_\gamma^{-1} \simeq 137.04$, subject to the stated hypotheses of \cite{gaugecouplings}; a correlated neutrino-cosmology package headed by the inverted mass ordering, to be decided by JUNO, with leptonic CP conservation and a computed quark CP phase \cite{teliCKM, teliLepCP}; hadronic electric dipole moments near current bounds, the generic residue of a parity-type strong-CP mechanism \cite{strongcp}; sterile-neutrino signatures, one per generation; dark-sector electromagnetism with kinetic-mixing phenomenology \cite{dark1, dark2, darkexp, finster}; and tests of the predicted supra-quantum (beyond-Tsirelson) correlations tied to the two-dimensional bridge of Section~\ref{sec:leaves} --- a pre-quantum residue expected only in the bridge-mediated regime there described, and absent in the leaf-local limit in which ordinary quantum theory is recovered, which is why existing Bell tests, which saturate but do not exceed the Tsirelson bound, are consistent with it \cite{Tsir, supra}. The registry \cite{catalogue} states, for each entry, the outcome that would refute it.

\section*{Acknowledgements}
We thank Matej Pav\v{s}i\v{c} for helpful correspondence and for bringing to our attention his early works on six-dimensional spacetimes, and Jose Isidro for insightful discussions and for drawing our attention to the literature on Yang--Mills descriptions of general relativity. We thank David Chester for a helpful observation on hypercharge normalisation, and the anonymous referee for questions on the anomaly structure, the relation to earlier work, and the experimental outlook, which shaped Sections~\ref{sec:anomalies}--\ref{sec:outlook}.

\medskip
\noindent{\bf Use of generative AI:} During the preparation of this manuscript, the authors used generative AI tools [Open AI's GPT-5.5 Pro and GPT-5.6 Sol, and Anthropic's Claude Fable 5] in adversarial mode for support in the technical analysis, organisation, writing, and editing of the manuscript. The original ideas are due to the authors, who take full intellectual responsibility for the content of the manuscript.

\end{document}